\documentclass[showpacs,twocolumn,floatfix,superscriptaddress]{revtex4}
\usepackage{graphicx}
\usepackage{amsmath}
\begin{document}

\title{Statistics of Heat Transfer in Mesoscopic Circuits}

\author{M. Kindermann\footnote{Present
address: Department of Physics, Massachusetts Institute of Technology,
Cambridge MA 02139, USA.}}
\affiliation{Instituut-Lorentz, Universiteit Leiden,
             P.O. Box 9506, 2300 RA Leiden, The Netherlands}
\author{S. Pilgram}
\affiliation{D\'epartement de Physique Th\'eorique, 
             Universit\'e de Gen\`eve,
             CH-1211 Gen\`eve 4, Switzerland}
\date{\today}

\begin{abstract}
A method to calculate the statistics of energy exchange between quantum
systems is presented. The generating function of this statistics is expressed
through a Keldysh path integral. The method is first applied to the problem of
heat dissipation from a biased mesoscopic conductor into the adjacent
reservoirs.  We then consider energy dissipation in an electrical  circuit
around a mesoscopic conductor. We derive the conditions under which
measurements of the fluctuations of heat  dissipation  can be used to
investigate higher order cumulants of the charge counting statistics of a
mesoscopic conductor.
\end{abstract}

\maketitle

During the second decade of mesoscopic physics there has been an increasing
interest in thermal phenomena. Several experiments succeeded in measuring
properties of heat transport in mesoscopic samples: Experimental tools based
on the Coulomb blockade to measure accurately local temperatures in mesoscopic
samples have been established \cite{Pekola1,Pekola2}.  It has been
demonstrated that the heat conductance through one-dimensional phonon modes of
a micro bridge is quantized at low temperatures \cite{Schwab1}. The universal
heat conductance per mode is given by $\pi^2k^2T/3h$.  The same universal heat
quantum is also found for carriers other than bosons (see \cite{Rego1} and
Refs. therein).  Thermopower was experimentally investigated for a quantum dot
in the Coulomb blockade regime \cite{Mol,Dzurak1} and for multiwalled carbon
nanotubes \cite{Kim1}.  The Andreev reflection process is ineffective for heat
transport and has been employed to measure thermopower in Andreev
interferometers. The thermopower depends on the magnetic flux and shows
geometry dependent time inversion properties \cite{Eom1}.

All these experiments have in common that they investigate mean properties
averaged over time. In electrical transport, however, noise measurements have
become a very useful tool to study properties of non-equilibrium systems which
are hidden in measurements of the mean current (for a review see
\cite{Blanter00}). A recent milestone was the first successful experimental
investigation of higher order current correlators by Reulet et
al. \cite{Reu03}.  It revealed an unexpected temperature dependence that was
explained by accounting for an external measurement circuit
\cite{Beenakker03}.  The calculation of average currents and current
fluctuations can be unified within the concept of full counting statistics
that has been introduced into mesoscopic physics a decade ago. It was shown by
Levitov and Lesovik that coherent charge transfer through a two-terminal
conductor can be seen as probabilistic process defined through a set of
transmission probabilities \cite{Lev93}.  Thereafter different works
investigated the statistics of other measurable quantities in mesoscopic
structures such as voltage \cite{Kindermann2}, momentum \cite{art5}, and
charge inside a mesoscopic volume \cite{Pilgram1}.

In this publication we address the statistics of fluctuations of energy
exchange.  We investigate the statistics of energy flow into a subvolume of a
quantum system.  In the first part of the paper, we write the generating
function of this statistics in a Keldysh-representation \cite{art8}.  The
``counting field'' that generates moments of transferred energy enters as a
gauge field that shifts the time evolution of the studied subvolume relative
to that of the rest of the system.

In section \ref{sc:tunnel} we apply our approach to a simple example which
nicely demonstrates the basic features of energy transfer: We consider the
energy dissipation from a biased mesoscopic conductor into one of its
reservoirs. The problem has been addressed before for a conductor at zero
temperature \cite{Ada01}.  In \cite{Rokni95} it was shown that dissipation
happens symmetrically, that is each reservoir dissipates half of the energy.
We find that this is true only on average and we determine fluctuations around
this rule. In contrast to the statistics of charge transfer through such a
conductor, the transfer of heat is not quantized.  In section \ref{sc:linear}
we derive general relations for the statistics of energy exchange between a
quantum system and a linear environment. Moments of the exchanged energy are
expressed in terms of correlators of the variable that couples the quantum
system to its environment. We apply these results to a mesoscopic conductor
which is embedded in a macroscopic measurement circuit in section
\ref{sc:circuit}. The energy transfer statistics is related to correlators of
the current through the mesoscopic conductor. One expects that the second
moment of energy dissipated in a series resistor depends on the fourth moment
of current fluctuations in the mesoscopic conductor. This suggests to use a
measurement of heat fluctuations as a probe to study higher order current
correlations in mesoscopic conductors. We quantify this expectation and derive
the conditions under which higher order current correlators can be extracted
from an energy measurement.

\section{Approach}
\begin{figure}

\includegraphics[width=5cm]{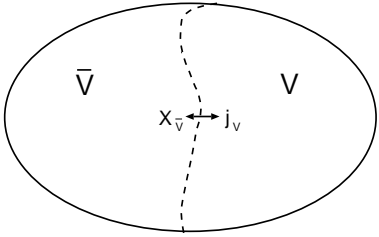}
\caption{ Quantum system divided into two subvolumes $V$ and $\bar{V}$ that
are coupled via the variables $j_V$ and $X_{\bar{V}}$.  }
\label{VbarV}

\end{figure}

We study the statistics of the exchange of energy between two subvolumes $V$
and $\bar{V}$ of a quantum system. We assume that no degrees of freedom leave
or enter $V$ and that $V$ is coupled to $\bar{V}$ only through one degree of
freedom $j_V$.  $j_V$ shall be coupled to the variable $X_{\bar{V}}$ in
$\bar{V}$, see Fig.\ \ref{VbarV}. A generalization of our approach to more
than one coupling variable is straightforward. Our model Hamiltonian reads
\begin{equation}
 H = H_V + H_{\bar{V}} + j_V X_{\bar{V}}
 \end{equation}
with  $[j_V,X_{\bar{V}}]=[j_V, H_{\bar{V}}]=[H_V, X_{\bar{V}}]=[H_V, H_{\bar{V}}]=0$. 
 
To calculate the statistics of the flow of energy into $V$ in a period of time
$[0,\tau]$ we assume it to be in an eigenstate $|E_V\rangle$ of $H_V$ at time
$t=0$ and calculate moments of $H_V$ at time $t=\tau$, defining the generating
function
\begin{equation} \label{eq:heat}
 \bar{\cal Z}(\xi)= \langle e^{i \xi H_V/2} e^{i H \tau} e^{-i \xi H_V}e^{-i  H \tau} e^{i \xi H_V/2} \rangle.
\end{equation}
(We have set $\hbar=1$.)  The average is taken over the initial state
$|E_V\rangle$.  The exponential $\exp\{ -i \xi H_V\}$ generates moments of the
energy in $V$ at time $\tau$.  Through the exponentials $\exp\{i \xi H_V/2\}
|E_V\rangle = \exp\{i \xi E_V/2\} |E_V\rangle$ and $\langle E_V| \exp\{i \xi
H_V/2\} $ the initial energy $E_V$ in $V$ is subtracted from that such that
$\bar{\cal Z}$ generates moments of $\triangle H_V= e^{i H \tau} H_V e^{-i H
\tau}-E_V $, the flux of energy into $V$ during $[0,\tau]$,
 \begin{equation} \label{eq:genHV}
 \langle (\triangle H_V) ^p \rangle = i^{p}\frac{\partial^p}{\partial \xi^p}\bar{\cal Z}(\xi)\Big|_{\xi=0}.
 \end{equation}
A generalization of $\bar{\cal Z}$ that generates correlators of the energy
flux at finite frequency is the functional
  \begin{eqnarray} \label{eq:heatfreqorig}
 {\cal Z}[\vec{\xi}]&=& \left\langle \overleftarrow{T} e^{i\int{dt\, [H- \dot{\xi}^-(t) H_V]}} \right.\nonumber \\
 && \left. \;\; \overrightarrow{T} e^{-i\int{dt\,[H - \dot{\xi}^+(t) H_V]}}
\right\rangle.
 \end{eqnarray}
The symbols $\overrightarrow{T}(\overleftarrow{T})$ denote (inverse) time
ordering and we have collected the two source functions into a vector in a
``Keldysh space'', $\vec{\xi}=(\xi^+,\xi^-)$.  ${\cal Z}$ correlates the
energy flux at different times,
 \begin{equation} \label{eq:funcdiff}
 \left\langle \prod_{q=1}^p \dot{H}_V(t_{q})  \right\rangle = \left(\frac{i}{2}\right)^p \prod_{q=1}^p \left[\frac{\delta}{  \delta \xi^+(t_{q})}-\frac{\delta}{ \delta \xi^-(t_{q})}\right]  { {\cal Z}} \Big|_{{ {\xi} }=0}.
\end{equation}
${\cal Z}$ arises naturally as one models a linear detector that measures the
energy flux $\dot{H}_V$ into $V$ \cite{art8}. The detector read-off $r$ is
then a linear functional of $\dot{H}_V$,
  \begin{equation}
  \langle r (\tau)\rangle  = \left\langle \int{dt\, s(\tau-t) \dot{ H}_V(t)} \right\rangle,
 \end{equation} 
where the response function $s(t)$ is causal, $s(t)=0$ for $t<0$, and it
depends on the internal dynamics of the detector. Moments of the detector
read-off are then generated by ${\cal Z}$ according to
 \begin{equation} \label{eq:rn}
 \langle r^p(\tau) \rangle =  \left(\frac{i}{2}\right)^p \frac{\partial^p}{\partial \xi^p}{\cal Z}\left[\xi s(\tau-t)\choose -\xi s(\tau-t)\right]\Big|_{\xi=0}.
 \end{equation}
Evidently, Eq.\ (\ref{eq:heatfreqorig}) reduces to (\ref{eq:heat}) for the
measurement of the time integrated energy flow during $[0,\tau]$, if $s(t) =
\theta_{\tau}(t) \equiv \theta(t)-\theta(t-\tau)$ [with the step function
$\theta(t)=1$ for $t>0$ and $\theta(t)=0$ for $t \leq 0$].
 
In the interaction picture with respect to the uncoupled problem
$H_0=H_V+H_{\bar{V}}$, $j_V(t)= \exp\{i H_V t\} j_V \exp\{-i H_V t\}$ and $
X_{\bar{V}}(t)= \exp\{i H_{\bar{V}} t\} X_{\bar{V}} \exp\{-i H_{\bar{V}} t\}
$, ${\cal Z}$ takes the form
  \begin{eqnarray} \label{eq:heatfreq}
 {\cal Z}[\vec{\xi}]&=& \Big\langle \overleftarrow{T} e^{i\int{dt\, [X_{\bar{V}}(t) j_V(t) - \dot{\xi}^-(t) H_V]}} \nonumber \\
 && \overrightarrow{T} e^{-i\int{dt\,[X_{\bar{V}}(t) j_V(t) - \dot{\xi}^+(t) H_V]}}
\Big\rangle.
 \end{eqnarray}
 
We rewrite Eq.\ (\ref{eq:heatfreq}) by breaking the time-ordered product up
into a large number of time-development exponentials for infinitesimal time
steps $\epsilon$.  Applying to each one of them the identity
 \begin{equation}
  e^{-i H_V\xi(t)} e^{-i \epsilon j_V(t) X_{\bar{V}}(t)}  e^{i H_V \xi(t)} = e^{-i \epsilon j_V[t-\xi(t)]  X_{\bar{V}}(t)}
  \end{equation}
  we find that
 \begin{equation} \label{eq:shiftHam}
 {\cal Z}[\vec{\xi}]= \left\langle \overleftarrow{T} e^{i\int{dt\,  j_V[t-\xi^-(t)] X_{\bar{V}}(t)}}
 \overrightarrow{T} e^{-i\int{dt\,  j_V[t-\xi^+(t)] X_{\bar{V}}(t)}} \right\rangle.
 \end{equation}
The generating functional of the statistics of energy flow into a volume $V$
takes the form of a partition functional of the entire system with shifted
time arguments of all variables in $V$. An analogous structure has been found
before for the statistics of transfer of other globally conserved quantities,
like charge \cite{Lev93} and momentum \cite{art5}. Also there the generating
functional has the structure of Eq.\ (\ref{eq:shiftHam}). The source term
locally shifts the variable conjugated to the measured quantity (the phase for
a measurement of charge, position for a momentum measurement, and time for the
energy measurement considered here).

In a path integral formulation of Eq.\ (\ref{eq:heatfreqorig}) we mark fields
that evolve the system forward and backward in time with superscripts $+$ and
$-$ respectively and we collect them in vectors like the source $\vec{\xi}$.
We may change integration variables as $j^{\pm}_V(t-\xi^{\pm}(t))\to
j^{\pm}_V(t) $ in the action corresponding to $H_V$ if $|\dot{\xi}(t)| < 1$ at
all times (only then this map is bijective).  The path integral takes then a
form very similar to Eq.\ (\ref{eq:shiftHam}),
\begin{eqnarray} \label{eq:genpath}
 {\cal Z}[\vec{\xi}]&=& \int{ {\cal D} \vec{j}_V {\cal D} \vec{X}_{\bar{V}} \,e^{-i\left( {\cal S}_V[\vec{j}_V] + {\cal S}_{\bar{V}}[\vec{X}_{\bar{V}}]\right)}} \nonumber \\
&& \;\;\;\;\;\;\;\;\;\;\;\;\;\;\;\;\;\;\; \times e^{ -i\int{dt\,  \vec{j}_V(t-\vec{\xi}(t)) \tau_3 \vec{X}_{\bar{V}}(t)}}.
 \end{eqnarray}
We have introduced actions ${\cal S}_V$ and ${\cal S}_{\bar{V}}$ for the
uncoupled volumes $V$ and $\bar{V}$ that are obtained by integrating the
actions corresponding to $H_V$ and $H_{\bar{V}}$ over all variables except
$j_V$ and $X_{\bar{V}}$ respectively. $\tau_3$ is the third Pauli matrix and
we use the notation
$[\vec{j}_V(t-\vec{\xi}(t))]^{\alpha}={j}^{\alpha}_V(t-{\xi}^{\alpha}(t))$
($\alpha = \{+,-\}$). The assumption $|\dot{\xi}(t)| \leq 1$ has the following
physical interpretation. From Eq.\ (\ref{eq:genHV}) and its finite frequency
generalization it is seen that one can infer a probability distribution of
energy flow by Fourier transforming the generating function ${\cal
Z}(\xi)$. To resolve energy differences of the order of $\triangle E$ in this
distribution one needs to know ${\cal Z}$ for values $\xi_{\triangle E} \simeq
1/\triangle E$. The condition $|\dot{\xi}(t)| < 1$ means then that the energy
detector smears out the measurement over a time $\triangle t > \xi_{\triangle
E} \simeq 1/ \triangle E$, that is, it does not attempt to measure the energy
faster than it is allowed to by the uncertainty relation.

\section{Heat dissipation statistics in tunnel junctions} \label{sc:tunnel}

\begin{figure}[t]
\includegraphics[width=5cm]{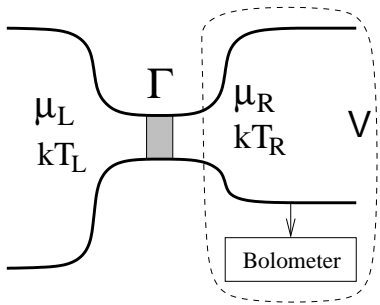}
\caption{Geometry considered in this section: A bolometer is attached to one
reservoir of a biased mesoscopic conductor. We calculate the statistics of
heat dissipated into the bolometer.}
\label{Geometry Of Contact}
\end{figure}
We first illustrate  our method with a particularly simple example: we
calculate the statistics of heat dissipated from a voltage biased mesoscopic
conductor into a bolometer attached to its right contact (see
Fig.~\ref{Geometry Of Contact}).  To this end, we express the action of a
simple connector in circuit theory by the Keldysh Green's functions $ 
G_{L,R}$ of the adjacent reservoirs \cite{Wolfgang,NazBag}
\begin{equation}
 S_{\rm con} = \frac{i}{2}\sum_{n} 
{\rm Tr} \ln\left[1+ \frac{1}{4}\Gamma_n
(\{  G_L,  G_R^\xi\} - 2)\right]
\label{Connector}
\end{equation}
where the $\Gamma_n$ denote energy independent transmission probabilities and
the brackets $\{,\}$ anticommutation.  The trace includes integration over
time.  In the following, we address the zero frequency limit only. In this
case, the time integration in Eq. (\ref{Connector}) is conveniently rewritten
as energy integration.  As formulated in Eq.\ (\ref{eq:shiftHam}), we obtain
the statistics of heat dissipation into the right contact by shifting all
observables of the right contact in time by $\xi_+$ on the upper Keldysh
contour and by $\xi_-$ on the lower contour.  The diagonal elements of $ G_R$
in the Keldysh space are left unchanged by this transformation which can be
cast in the following form (we introduce the difference $\xi=\xi_+-\xi_-$):
\begin{equation}
\begin{array}{c}
  G_R^\xi =
e^{ i \xi (\epsilon-\mu_R) \tau_3/2 }
  G^0_R(\epsilon)
e^{ -i \xi (\epsilon-\mu_R) \tau_3/2},\\
\quad\\
  G_j^0(\epsilon) = \left(
\begin{array}{cc}
 1-2f_j & 2 f_j \\
 2(1-f_j) & 2f_j-1
\end{array} \right).
\end{array}
\label{Greens Rotation}
\end{equation}
The time shift introduces a rotation of the Green's function $  G_R$
analogous to charge counting statistics \cite{Wolfgang}.  $f_j=f_j(\epsilon)$
denotes the energy dependent Fermi function of contact $j=L,R$ that depends on
temperature $T_j$ and electrochemical potential $\mu_j$.  Evaluating
Eq. (\ref{Connector}) in the zero frequency limit we find the cumulant
generating function \cite{Earlier1}
\begin{equation}
\label{Heat Statistics}
\begin{array}{c}
\ln {\cal Z}[\xi] =\\
\quad\\
 \frac{\tau}{2\pi} \int d\epsilon \sum_n
\ln\left\{1+\Gamma_nf_L(1-f_R)(e^{-i\xi(\epsilon-\mu_R)}-1)\right.\\
\quad\\
\left.+\Gamma_nf_R(1-f_L)(e^{i\xi(\epsilon-\mu_R)}-1) \right\}.
\end{array}
\end{equation}
Cumulants of dissipated energy can now be calculated easily by taking
derivatives with respect to $\xi$.  The mean dissipation without applied
voltage ($\mu_j=0$) is for instance given by
\begin{equation}
\langle \Delta H_V \rangle = 
\left.i\frac{\partial\ln{\cal Z}}{\partial \xi}\right|
_{\xi=0} =
\frac{\pi\tau}{12}\sum_n \Gamma_n
\left\{ \left(kT_L\right)^2 - \left(kT_R\right)^2
\right\}.
\end{equation}
For almost equal temperatures $T_L\rightarrow T_R$ we recover the mesoscopic
version of the Wiedemann-Franz law for thermal conductance
\cite{Imry86,Bruder97}.
\begin{figure}[t]
\includegraphics[width=8cm]{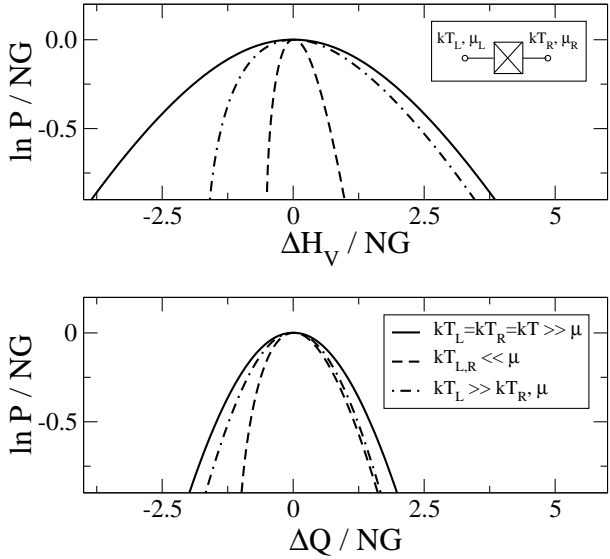}
\caption{Comparison of energy dissipation $\Delta H_V = H_V - \langle H_V
\rangle$ (upper panel) and charge $\Delta Q= Q - \langle Q \rangle$ statistics
(lower panel) for a tunnel contact in equilibrium and non-equilibrium due to
an applied bias or temperature difference.  The energy is normalized to
temperature $kT_L$ or bias $\mu = \mu_L - \mu_R$ respectively.  The product
of $N=\mu\tau/2\pi,kT_L\tau/2\pi$ and dimensionless conductance $G$ is
assumed to be large.  }
\label{Different Statistics}
\end{figure}

We note that Eq.\ (\ref{Heat Statistics}) is similar to the statistics of
charge transfer\cite{Lev93}: Charge statistics is recovered by substituting
$\xi(\epsilon-\mu_R) \mapsto \chi e$ where $e$ is the elementary charge and
the field $\chi$ generates cumulants of charge transfer.  As for charge
statistics, there is a classical probabilistic interpretation of Eq.\
(\ref{Heat Statistics}): An electron of total energy $\epsilon$ in channel $n$
passes the mesoscopic conductor with probability $\Gamma_n$. It then
dissipates the energy $\epsilon-\mu_R$ in the right contact.  It is the energy
integration in Eq.\ (\ref{Heat Statistics}) which makes the statistics of heat
transfer interesting.  Unlike in the  binomial charge statistics the exponent
$i\xi(\epsilon-\mu_R)$ assigns energy weights to each electron. Therefore,
energy transfer is not quantized.

We now turn to a specific example: the tunnel junction with $\Gamma_n\ll
1$. In this case, the logarithm of Eq.\ (\ref{Heat Statistics}) can be
expanded and the energy integration involves only elementary integrals (for
equal temperatures $T_L = T_R = T$ on both sides of the junction).  Defining
$\mu = \mu_L - \mu_R$ we find the following result for the characteristic
function:
\begin{equation}
\label{Characteristic Tunnel}
\begin{array}{c}
\ln {\cal Z}[\xi] = G\tau \coth\frac{\mu}{2kT}
\left(kT\frac{\sin(\xi\mu)}{\sinh(\pi\xi kT)/\pi}-\mu\right)\\
\\
-iG\tau \frac{1-\cos(\xi\mu)}{\sinh(\pi\xi kT)/\pi}.
\end{array}
\end{equation}
We introduced the conductance of the junction $G=\sum \Gamma_n /2\pi$.  For
comparison, we also give the statistics of charge transfer \cite{Levitov01}
\begin{equation}
\label{Characteristic Charge Tunnel}
\ln {\cal Z}[\chi] = G\tau\mu\coth\frac{\mu}{2kT}
\left(\cos(\chi e)-1\right) - iG\mu\sin(\chi e).
\end{equation}
We observe several differences: Whereas ${\cal Z}[\chi]$ is strictly periodic,
the characteristic function of heat dissipation ${\cal Z}[\xi]$ shows damped
oscillations.  The lacking periodicity is due to the unquantized transfer of
energy. For charge transfer, the real part of ${\cal Z}[\chi]$ is even in the
applied voltage $\mu$ in contrast to an odd imaginary part. Odd cumulants
therefore change sign under voltage inversion.  This is not the case for
energy, ${\cal Z}[\xi]$ does not depend on the sign of the applied voltage.
Heat dissipation takes place in both reservoirs symmetrically regardless of
the current direction! Odd cumulants of charge transfer (the imaginary part of
Eq.~(\ref{Characteristic Charge Tunnel})) do not depend on
temperature\cite{Levitov01}. In contrast, we find that the asymmetry of the
distribution of dissipated heat does depend on temperature (see the imaginary
part of Eq.~(\ref{Characteristic Tunnel})).

Another interesting case is the statistics in the presence of a
temperature gradient only. Analytical results are available for $T_L\gg
T_R$. We find
\begin{equation}
\label{Temperature Gradient}
\begin{array}{c}
\ln {\cal Z}[\xi] =\\
\quad\\
 2G\tau k T_L 
\left\{\frac{i\pi}{\sinh(\pi k T_L \xi)}
+\ln(2) - \beta(-ikT_L \xi) \right\}=\\
\quad\\
G\tau kT_L \left\{-i\frac{\pi^2}{6}kT_L\xi - \frac{3\zeta(3)}{2}
(kT_L\xi)^2
+i\frac{7\pi^4}{360}(kT_L\xi)^3 % + \frac{15\zeta(5)}{8}\xi^4 
+ .. \right\}
\end{array}
\end{equation}
for energy transport and
\begin{equation}
\label{Temperature Gradient Charge}
\ln {\cal Z}[\chi] = 2G\tau k T_L \ln(2) 
\left(\cos(\chi e)-1\right)
\end{equation}
for charge transport (we introduced the $\beta$-function $\beta(iz) =
\sum_{n=0}^{\infty} (-1)^n / (iz + n)$). Fig.\ \ref{Different Statistics}
illustrates several limiting cases: On a log scale, it compares the charge and
energy transfer statistics in equilibrium, and in non-equilibrium due to an
applied voltage (see Eqs.\ (\ref{Characteristic Tunnel}) and
(\ref{Characteristic Charge Tunnel})) and due to an applied temperature
gradient (see Eqs.\ (\ref{Temperature Gradient}) and (\ref{Temperature
Gradient Charge})).
\begin{figure}[t]
\includegraphics[width=8cm]{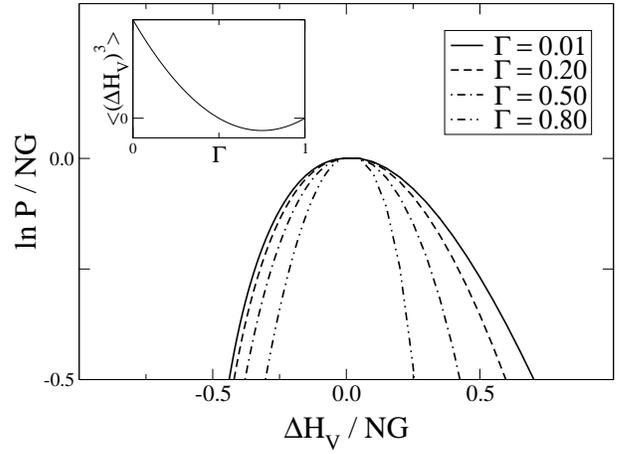}
\caption{Probability Distribution of dissipated energy into one reservoir of a
biased mesoscopic conductor ($\mu \gg kT$).  The distribution is broader for
tunneling junctions ($\Gamma=0.01$) than for open point contacts
($\Gamma=0.80$). It is clearly seen that the third cumulant of the
distribution changes sign.  The inset shows the third cumulant $\langle
(\Delta H_V)^3 \rangle$ as a function of the transparency $\Gamma$.  ($G$ is
the dimensionless conductance and $N=\mu\tau/2\pi$, we assume $NG$ to be
large)}
\label{Different Transparencies}
\end{figure}

In general, Eq.\ (\ref{Heat Statistics}) has to be evaluated
numerically. Fig.\ \ref{Different Transparencies} shows the probability
distribution of dissipated heat for barriers with various transparencies
$\Gamma_n = \Gamma$ in the high-voltage regime $\mu\gg kT$. Noise disappears
in the ballistic limit $\Gamma=1$. It is clearly visible that the third
cumulant of the distribution changes its sign as a function of $\Gamma$. We
find that
\begin{equation}
\label{Skewness}
\langle (\Delta H_V)^3 \rangle = G \mu^4 (1-\Gamma)(1-2\Gamma) / 4
\end{equation}
(see also the inset of Fig.\ \ref{Different Transparencies}).

In this entire section, we assumed that the energy dissipation in the right
reservoir is measured by an ideal bolometer which does not act back on the
mesoscopic contact. A realistic bolometer is characterized by a finite
thermal conductance.  Energy fluctuations in the right reservoir are then
converted into temperature fluctuations which modulate the noise intensity of
the mesoscopic contact. This backaction is similar to the backaction from a
non-ideal current meter\cite{Beenakker03}.

\section{Energy flow into a linear medium} \label{sc:linear}

We use now Eq.\ (\ref{eq:genpath}) to calculate the statistics of energy flow
from a system $S$ into a linear medium. An example is the energy that $S$
emits as electromagnetic radiation. The electromagnetic field is then the
linear medium into which energy flows. This energy flow may for example be
measured with a photo-detector.  The distribution of the number of photons
emitted by a source current $j_S$ can be expressed in terms of correlators of
$j_S$ \cite{Fle98}. In Ref.\ \cite{Fle98} this relation has been established
perturbatively in a weak coupling of the photo-detector to the electromagnetic
field. In Ref.\ \cite{Bee00} it has been used to calculate the fluctuations in
the number of photons emitted by a mesoscopic conductor.  Eq.\
(\ref{eq:genpath}) allows us to obtain these results non-perturbatively. As a
second application of our method we calculate the fluctuations in the amount
of heat that is dissipated in a macroscopic electrical circuit around a
mesoscopic conductor. Such fluctuations could be measured with a bolometer.
  
\subsection{General relations}
\begin{figure}

\includegraphics[width=8cm]{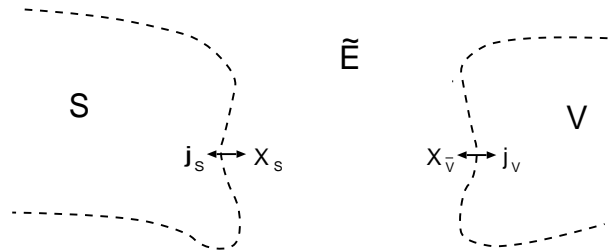}
\caption{ $S$ is an arbitrary quantum system. The energy flow into the part
$V$ of its linear environment is considered in this section. The variables
$j_S$, $X_S$ and $j_V$, $X_{\bar{V}}$ couple the three constituent volumes.  }
\label{couplefig}

\end{figure}

In our analysis we divide space into three regions. The system $S$, a volume
$V$ into which the energy flow is measured and a region $\tilde{E}$. We call
$E=\tilde{E}+V$ the environment of $S$. The subspaces are connected via two
variables $X_S$ and $X_{\bar{V}}$ in $\tilde{E}$, that couple to $j_S$ in $S$
and $j_V$ in $V$ respectively (see Fig.\ \ref{couplefig}). With the
Hamiltonians $H_S$, $H_{\tilde{E}}$, and $H_V$ of $S$, $\tilde{E}$, and $V$ we
then have the total Hamiltonian
\begin{equation}
H= H_S + H_{\tilde{E}} + H_V + X_S (j_S-\bar{j}_S) + X_{\bar{V}}
(j_V-\bar{j}_V),
\end{equation}
where $\bar{j}_V$ and $\bar{j}_S$ are sources that generate moments of
$X_{\bar{V}}$ and $X_S$.
The generating functional for correlators of the energy flow into $V$ takes
the form of Eq.\ (\ref{eq:genpath}) with
\begin{eqnarray} 
 e^{-i {\cal S}_{\bar{V}}[\vec{X}_{\bar{V}}] }&=& \int{ {\cal D} \vec{X}_S {\cal D} \vec{j}_{S} \, e^{-i \left({\cal S}_S[\vec{j}_S] + {\cal S}_{\tilde{E}}[\vec{X}_{S},\vec{X}_{\bar{V}}]\right)}} \nonumber \\
 && \;\;\;\;\;\;\;\;\;\;\;\;\;\;\;\; \times e^{ -i \int{dt\,  \vec{j}_S(t) \tau_3 \vec{X}_{S}(t)}}
 \end{eqnarray}
with the actions ${\cal S}_S$ and ${\cal S}_{\tilde{E}}$ corresponding to
$H_S$ and $H_{\tilde{E}}$.  $ {\cal S}_{\tilde{E}}$, being the action of a
linear system, is quadratic and depends only on the response functions of
$\tilde{E}$ and its temperature $T$. We characterize the response of $E$ when
disconnected from $S$ (corresponding to the Hamiltonian $H-H_S-X_S j_S$) by
four functions,
\begin{eqnarray}
R_{SS}(\omega)&=&\frac{\partial X_S(\omega)}{\partial \bar{j}_S(\omega)} \Big|_{\bar{j}_V} ,\;\;\;
R_{SV}(\omega)=\frac{\partial X_S(\omega)}{\partial \bar{j}_V(\omega)} \Big|_{\bar{j}_S} , \nonumber \\
R_{VS}(\omega)&=&\frac{\partial X_{\bar{V}}(\omega)}{\partial \bar{j}_S(\omega)} \Big|_{\bar{j}_V} ,\;\;\;
R_{VV}(\omega)=\frac{\partial X_{\bar{V}}(\omega)}{\partial \bar{j}_V(\omega)} \Big|_{\bar{j}_S}. \nonumber \\
\end{eqnarray}
The volume $V$ is described by the response to a source $\bar{X}_V$ coupling
to $j_V$ in the absence of $\tilde{E}$, corresponding to the Hamiltonian $H_V
- j_V \bar{X}_{{\bar{V}}}$,
\begin{equation}
R^{-1}_{iso}(\omega)=\frac{\partial j_V(\omega)}{\partial \bar{X}_{{\bar{V}}}(\omega)} . 
\end{equation}
The environment's action is determined by the fluctuation dissipation theorem
\cite{art8}.  The action for $j_V$ when $V$ is isolated reads
\begin{equation}
{\cal S}_{iso}[\vec{j}_V]= \vec{j}_V\otimes M_{iso} \otimes   \vec{j}_V.
\end{equation}
The matrix multiplication $\otimes$ extends over Keldysh as well as time
indices, $[A\otimes B](t,t'') \equiv \int{dt'\,A(t,t') B(t',t'')}$ and
correspondingly for vectors like $\vec{j}_V(t)$.  In a steady state
$M_{iso}(t,t')$ depends only on the time difference $t-t'$ and in Fourier
representation $M(\omega) = \int{d(t-t') \, e^{i\omega (t-t')} M(t-t')}$ it
reads
\begin{eqnarray}
M_{iso}(\omega)&=& {\textstyle \frac{1}{2}} R_{iso}(\omega) \tau_3 + i \,{\rm Im}\, R_{iso}(\omega)  G_0(\omega), \\
G_0(\omega)&=& \left( \begin{array}{cc}
N(\omega) & -N(\omega)  \\ -N(\omega)-1  &  N(\omega)+1 \end{array} \right),
\end{eqnarray}
with the Bose-Einstein distribution $N(\omega)=(\exp\{ \omega/k_B
T\}-1)^{-1}$. (We define the response functions such that they have negative
imaginary part.)
 
We analyze first the particularly simple case that the energy flow into the
entire environment to $S$ is measured, $V=E$. Then we have
${X}_{\bar{V}}=j_S$, $X_S=j_V$ and $ {\cal S}_{\bar{V}} = {\cal S}_S$. We
rewrite Eq.\ (\ref{eq:genpath}) by introducing a coupling matrix
\begin{equation}
\sigma(t,t') = \left( \begin{array}{cc}
\delta\biglb(t-t'+\xi^+(t')\bigrb) &   \\
0 & \delta\biglb(t-t'+\xi^-(t')\bigrb)    
\end{array} \right),
\end{equation}
\begin{equation} \label{eq:VisE}
 {\cal Z}[\vec{\xi}]= \int{ {\cal D} \vec{j}_S {\cal D} \vec{j}_{V} e^{-i\left( {\cal S}_S[\vec{j}_S] +{\cal S}_{iso}[\vec{j}_V]+  \vec{j}_V \otimes \tau_3  \sigma \otimes \vec{j}_{S}\right)}}.
 \end{equation}
The Gaussian integrals over $\vec{j}_V$ in (\ref{eq:VisE}) are easily done,
resulting in
\begin{equation} \label{eq:Zxi}
 {\cal Z}[\vec{\xi}]= \int{ {\cal D} \vec{j}_S\, e^{-i\left( {\cal S}_S[\vec{j}_S] + {\cal S}_{E}[\vec{j}_{{S}}]  + {\cal A}_{\vec{\xi}}[\vec{j}_S]\right) }}
 \end{equation}
with an action $ {\cal S}_E$ that describes the influence of the environment
on $S$,
 \begin{equation} \label{eq:Senv}
  {\cal S}_{E}[\vec{j}_{{S}}]=  -{\textstyle \frac{1}{4}} \vec{j}_S\tau_3\otimes M_{iso}^{-1} \otimes  \tau_3 \vec{j}_S,
  \end{equation}
and a source term $ {\cal A}_{\vec{\xi}}$ that vanishes at $\vec{\xi}=0$,
   \begin{eqnarray} \label{eq:Axi}
  {\cal A}_{\vec{\xi}}[\vec{j}_S]&=&  \vec{j}_S\otimes A_{\vec{\xi}} \otimes  \vec{j}_S  , \;
  A_{\vec{\xi}}=  -{\textstyle \frac{1}{4}}   \tau_3 a_{\vec{\xi}}   \tau_3, \label{eq:Adirect}  \\
  a_{\vec{\xi}}&=&\sigma^{\dagger} \otimes M_{iso}^{-1} \otimes \sigma - M_{iso}^{-1}. 
  \end{eqnarray}
In the general case $V \neq E$, the environment's action is determined by
$R_{SS}$,
\begin{eqnarray}
  {\cal S}_{E}[\vec{j}_{S}]&=&  \vec{j}_S \otimes G_{SS} \otimes   \vec{j}_S,\\
 G_{\alpha\gamma}&=&  {\textstyle \frac{1}{2}} [R_{\alpha\gamma} \tau_3 +  (R_{\alpha\gamma}-R_{\gamma\alpha}^*) G_0],\;\; \; \alpha,\gamma \in \{S,V\}. \nonumber \\
\end{eqnarray}
The source term then takes the form
\begin{widetext}
 \begin{eqnarray} 
  A_{\vec{\xi}}&=&   G_{SV}\otimes\left\{[G_{VV}+G_{VV}\otimes a_{\vec{\xi}} \otimes G_{VV})]^{-1}-G_{VV}^{-1} \right\} 
  \otimes G_{VS}.  \nonumber \\
  \end{eqnarray}
  \end{widetext}
Eq.\ (\ref{eq:Adirect}) is recovered for $X_{\bar{V}}=j_S$ and $ X_S=j_V$,
such that $R_{VV}=0$, $R_{SS}=-R_{\rm iso}^{-1}$, and $R_{VS}=R_{SV}=1$.
 
\subsection{Zero frequency}

For concrete results we focus on zero frequency correlators of the energy
flow, choosing $\vec{\xi} = (1,-1) \theta_{\tau}(t)\xi /2$ [c.f. Eq.\
(\ref{eq:rn})].  We neglect transient effects, that is, we calculate only
terms of leading order in the time $\tau$ over which energy is accumulated. We
may then work in the discrete Fourier space of $\tau$-periodic functions $f$,
with Fourier coefficients $f_l= \int_0^{\tau}{dt\, e^{i\omega_l t} \,
f(t)}/\tau$ and frequencies $\omega_l = 2\pi l/\tau$. In this representation
the source term in Eq.\ (\ref{eq:Zxi}) reads
   \begin{equation} \label{eq:diagonal}
     {\cal A}_{\xi}[\vec{j}_S]= \tau \sum_l \vec{j}_{S,l} A_{\xi,l} \vec{j}_{S,l} .  
     \end{equation}
$\sigma$ is now diagonal in frequency indices,
\begin{equation}
\sigma_l = \left(\begin{array}{cc} e^{-i \omega_l\xi/2} & 0 \\ 0 & e^{i \omega_l\xi/2} \end{array} \right),
\end{equation} 
and so is $A_{\xi}$.  At zero temperature it takes a simple form even in the
general case $E \neq V$,
     \begin{equation} \label{eq:zerotemp}
   A_{\xi,l} =    |R_{VS}(\omega_l)|^2 \left( \begin{array}{cc}
0 & 0\\ i {\rm Im}\, R_{iso}^{-1}(\omega_l) \left(e^{-i \xi \omega_l}-1\right)  &  0 \end{array} \right).
\end{equation}
Applied to a linear photo-detector this reproduces the distribution of
photo-counts in response to a source $j_S$ obtained perturbatively in
\cite{Fle98}.  To see this we substitute $\xi \omega_l \to \xi_n$ in Eqs.\
(\ref{eq:Zxi}) with (\ref{eq:diagonal}) and (\ref{eq:zerotemp}) to obtain the
statistics of the number of absorbed quanta $n$ rather than that of the
absorbed energy. The further substitution $ e^{-i\xi_n}-1 \to \chi_n$ yields
the generating function for factorial moments of $n$. To compare the resulting
factorial moments of the photocount to the formulas obtained in \cite{Fle98},
we write them in the time domain,
\begin{widetext}
\begin{eqnarray} \label{eq:factorial}
\langle n(n-1)\cdots (n-p) \rangle = \Big\langle \Big[ \int_{-\infty}^{\infty}{ dt'\, dt'' \int_0^{\tau}{d\tau' \, d\tau'' }} 
 \,{\rm Im} \,R^{-1}_{iso}(t',t'') R_{VS}(t',\tau') R_{VS}(t'',\tau'') j_S^-(\tau') j_S^+(\tau'') \Big]^p \Big\rangle_{{\cal S}_S+{\cal S}_E}. \nonumber \\ 
\end{eqnarray}
\end{widetext}
This is equivalent to Eq.\ (30) in \cite{Fle98}. The detector sensitivity $f$
there corresponds to our ${\rm Im} \,R^{-1}_{iso}$, the density of absorbing
detector modes per unit of frequency. The retarded photon propagator $D_{ret}$
corresponds to our cross-response function $i R_{VS}$.  The time-ordering of
$j_S^+$ and $j_S^-$ along the Keldysh contour corresponds to the ``apex'' time
order of the sources in \cite{Fle98}.  The expectation value in Eq.\
(\ref{eq:factorial}) is taken with respect to the action $S_S$ of the source
system and the piece $S_E$ that describes the back action of the environment
on $S$.

\section{Linear electrical circuits} \label{sc:circuit}

Electrical conductors couple to their electromagnetic environment - the
circuit that they are embedded in - with the current $I_S$ that flows through
them.  To calculate the flow of energy into this environment we may therefore
apply the formulas obtained in the previous section with $j_S=I_S$, $H_S$
being the Hamiltonian of the conductor.  Also in this case energy is
transferred by means of photons. These photons may be either detected by a
photocounter \cite{Bee00} or with a bolometer, that measures the thermal
energy exchanged by means of photons.  According to Eq.\ (\ref{eq:factorial})
the $n$-th factorial moment of energy transfer is proportional to the $2n$-th
moment of fluctuations of the source $j_S$. This suggests that a measurement
of the energy flow and its fluctuations may be a useful tool for measuring
higher order correlators of electrical currents whose measurement poses an
experimental challenge. While experimental techniques for the measurement of
the variance of electrical current fluctuations are by now well developed, so
far only one experiment has been successful in measuring higher order current
correlators. In \cite{Reu03} the measurement of the third moment of current
fluctuations through a tunnel barrier has been reported. Eq.\
(\ref{eq:factorial}) suggests that the fourth moment of current fluctuations
can be inferred from the variance of the heat produced by them.

\begin{figure}

\includegraphics[width=8.5cm]{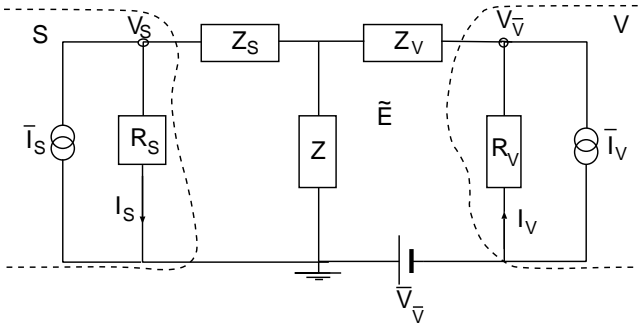}
\caption{Electrical circuit analyzed in this section. $R_S$ is the conductor
that generates current fluctuations $\delta I_S$. These fluctuations can be
characterized by a measurement of fluctuations of the heat that is dissipated
in the external resistor $R_V$.  }
\label{fig2}

\end{figure}

To quantify this we apply now the relations obtained in the previous section
to the electrical circuit depicted in Fig.\ \ref{fig2}. We assume all
resistors in the circuit to be macroscopic and linear, except $R_S$, that
plays now the role of the system $S$.  Fluctuations of the amount of heat that
is dissipated in $R_V$ as measured with a bolometer are by virtue of Eq.\
(\ref{eq:factorial}) related to fluctuations in the electrical current through
$R_S$.  For that the volume $V$ should be chosen the resistor $R_V$.  The
current $I_V$ flowing through the volume $V$ is coupled to the variable
$\chi_{\bar{V}}$ in the rest of the environment $\tilde{E}$.  This variable
$\chi_{\bar{V}}$ is the time integral of the voltage $V_{\bar{V}}$ over $R_V$,
$\chi_{\bar{V}}(t)=\int^t{dt'\,V_{\bar{V}}(t')}$. The current $I_S$ through
$R_S$ similarly couples to the environment variable
$\chi_S(t)=\int^t{dt'\,V_S(t')}$. Again we introduce external sources
$\bar{I}_V$ and $\bar{\chi}_{\bar{V}}$. The circuit Fig.\ \ref{fig2} is then
mapped onto the general model of Fig.\ \ref{couplefig} with the choice
$X_{\bar{V}}= \chi_{\bar{V}}$, $\bar{X}_{\bar{V}}= \bar{\chi}_{\bar{V}}$,
$X_S= \chi_S$, $j_V=I_V$, $\bar{j}_V= \bar{I}_V$, $ \bar{j}_S=\bar{I}_S$ and
$j_S=I_S$.  We need the response functions of the circuit Fig. \ref{fig2}
without the conductor $R_S$,
\begin{eqnarray}
R_{SS}&=&\frac{\partial \chi_S}{\partial \bar{I}_S} \Big|_{\bar{I}_V} =
\frac{1}{i\omega}\left[ Z_S +  \left( \frac{1}{Z}+\frac{1}{Z_V+R_V}
  \right)^{-1} \right] ,\nonumber
\end{eqnarray}
\begin{eqnarray}
R_{VS}&=&R_{SV}=\frac{\partial \chi_S}{\partial \bar{I}_V} \Big|_{\bar{I}_S} =
\frac{1}{i\omega}\frac{Z R_V}{Z+Z_V+R_V},
\end{eqnarray}
\begin{eqnarray}
 R_{VV}&=&\frac{\partial \chi_{\bar{V}}}{\partial \bar{I}_V} \Big|_{\bar{I}_S} =
 \frac{1}{i\omega}\left( \frac{1}{R_V}+\frac{1}{Z_V+Z} \right)^{-1}
\end{eqnarray}
\begin{eqnarray}
 R^{-1}_{iso}&=&\frac{\partial I_V}{\partial \bar{\chi}_{\bar{V}}} \Big|_{\rm isolated} = \frac{i\omega}{R_V} 
\end{eqnarray}
(all quantities here are frequency-dependent).  Note, that any four-terminal
circuit connecting the conductor $R_S$ with $R_V$ can be modeled by the three
resistors $Z$, $Z_S$, and $Z_V$.  The resulting zero-frequency energy flow
statistics takes a particularly simple form in the limit of an infinite $Z
$. Then the current flowing through $R_S$ is directly fed into $R_V$ and the
source term in the generating functional (\ref{eq:Zxi}) is given by Eq.\
(\ref{eq:diagonal}) with
\begin{eqnarray} \label{eq:Zinf}
   A_{\xi,l}& =&   - \frac{{\rm Re}\, R_V(\omega_l)}{i\omega_l}  \nonumber \\
   &&\!\!\!\!\!\!\!\! \!\!\!\!\!\!\!\! \!\!\!\!\!\!\!\!\!\!\!\!\!\!\! \times \left( \begin{array}{cc}
0 &  N(\omega_l) \left( e^{i \xi \omega_l}-1\right)  \\  \left[N(\omega_l)+1\right] \left( e^{-i \xi \omega_l}-1\right)  &  0 \end{array} \right) .
\end{eqnarray}
This is a generalization of Eq.\ (\ref{eq:zerotemp}) to finite temperature,
allowing for emission of energy quanta from $V$ as well as absorption. The
transferred quanta have predominantly energies that are smaller than the
inverse of the RC-time $\tau_{RC}$ of $R_V$. For finite $Z$ thermal current
fluctuations $\langle \delta I^2\rangle_{th} \approx kT/Z$ created in $Z$ mix
into the fluctuations of $I_S$. As a consequence Eq.\ (\ref{eq:Zinf}) is then
only valid in the regime $(\tau_{RC}kT)^2 \ll |Z/R_V|$.

The formulas for the statistics when the mixing in of thermal fluctuations
occurs are more complicated. To make further progress we assume that the
frequency dispersion of the measured current correlators is negligible on the
scale $1/\tau_{RC}$, $\langle \langle \prod_{q=1}^p I_S(\omega_q) \rangle
\rangle \approx 2\pi \delta(\sum_{q=1}^p \omega_q) C_p$ for $\omega_q \ll
1/\tau_{RC}$ (here $\langle \langle \cdots \rangle \rangle$ denotes
irreducible, or cumulant correlators). For a mesoscopic conductor this is
satisfied if $1/\tau_{RC}$ is smaller than the voltage applied to the
conductor. In this limit we find that
\begin{widetext}
\begin{eqnarray}
\langle \triangle H_V \rangle &=&   2\tau \int_0^{\infty}{\frac{d\omega}{2\pi}\,  R_{eff}(\omega)  C_2}  \label{eq:HV1} \\
\langle\langle  (\triangle H_V)^2 \rangle\rangle &=& 4  \tau \left\{ \left[  \int_0^{\infty}{\frac{d\omega}{2\pi}\, R_{eff}(\omega) }\right]^2 C_4   +   \int_0^{\infty}{\frac{d\omega}{2\pi}} \, \left[   R^2_{eff}(\omega) C^2_2  - \frac{\omega}{2} \biglb(2N(\omega)+1\bigrb) R_{eff}(\omega) C_2 \right] \right\}  \label{eq:HV2} 
\end{eqnarray}
%\end{widetext}
with $R_{eff}= {\rm Re}\, R_V |Z/(Z+Z_V+R_V)|^2$.  We have assumed that
$R_{eff}(0)=0$ and $R_{eff}(\omega) \to 0$ for $\omega > 1/\tau_{RC}$ such
that the DC-component (that will be avoided in experiments) and the
contributions from frequencies $\omega > 1/\tau_{RC}$ in Eqs.\ (\ref{eq:HV1})
and (\ref{eq:HV2}) are negligible .  We conclude that the variance of the heat
dissipated in $R_V$ depends on the fourth cumulant of current fluctuations in
the mesoscopic conductor, as one would expect. There are, however, also
contributions from lower order current correlators. The environment circuit
has a finite response time $\tau_{RC}$ and therefore effectively averages
fluctuations over that time.  As a result, the energy fluctuations become
dominated by the lowest order current correlator in the limit of a long
$\tau_{RC}$, when higher order current correlators become negligible and the
statistics becomes Gaussian. In this limit of ``narrow-band detection'' the
statistics of energy transfer is negative binomial \cite{Bee00}. Only
deviations from this encode non-Gaussian current correlations. In order to see
them the statistics of charge flow through the conductor during $\tau_{RC}$
has to be strongly non-Gaussian. This is the case if $\tau_{RC} \bar{I}/e$,
the mean number of transmitted electrons in that period, is small.
  
 More concretely, for a measurement of the fourth cumulant $C_4$ of current
fluctuations in $R_S$ one would want the first term in Eq.\ (\ref{eq:HV2}) to
be dominant.  An estimate of the three summands in Eq.\ (\ref{eq:HV2}) for a
tunnel contact with mean current $\bar{I}$ (assuming that $R_{eff}(\omega)
\approx R_{eff} $ for $\omega < 1/\tau_{RC}$),
%\begin{widetext}
\begin{eqnarray}
\langle\langle  (\triangle H_V)^2 \rangle\rangle \approx  \frac{\tau}{\tau_{RC}^2}  R_{eff}^2 G_Q e \bar{I} \left( 1 + \frac{\tau_{RC} \bar{I}}{e} + \frac{\tau_{RC}^2 kT\, {\rm min}\{kT,1/\tau_{RC}\} }{G_Q R_{eff}} \right)
\end{eqnarray}
\end{widetext}
(with the conductance quantum $G_Q=e^2/2\pi$) confirms this. The variance of
fluctuations of the heat flux into $R_V$ is a direct measure for the fourth
cumulant of current fluctuations in a tunnel contact if $\tau_{RC} \bar{I}/e
\ll 1$ and $\tau_{RC} kT \ll G_Q R_{eff} $. The origin of the first condition
has been explained qualitatively above. The second condition ensures that the
fourth cumulant is visible on the thermal background.  Back action effects of
the measuring resistor on the measured tunnel contact (conductance $G$) are
avoided if additionally $R_{SS} G \ll 1$ \cite{art8}.  These requirements are
rather restrictive, but can in principle be met in experiments. For a
practical implementation, the first condition $\tau_{RC} \bar{I}/e \ll 1$ may
be relaxed to $\tau_{RC} \bar{I}/e \simeq 1$ by measuring the voltage
dependence of $\langle\langle (\triangle H_V)^2 \rangle\rangle$ and extracting
the term that is linear in $\bar{I}$.

\section{Conclusion}

We have presented a theory for the statistics of energy exchange between
coupled quantum systems. As an application we have calculated the statistics
of energy dissipation into the leads connected to a mesoscopic
conductor. General and exact expressions can be obtained for the energy flow
from a quantum system into a linear environment. We have used these results to
calculate the moments of the heat dissipated in a linear circuit around a
mesoscopic conductor. They have been expressed in terms of moments of the
current fluctuations produced by the conductor. As one expects, the variance
of this dissipated heat depends on the fourth cumulative moment of current
fluctuations produced by the conductor. Heat fluctuations may therefore serve
as a tool to detect the fourth cumulant, whose direct measurement is
difficult. We have analyzed the conditions under which a measurement of heat
fluctuations can reveal higher order cumulants of the charge counting
statistics.

\begin{acknowledgements}
We thank Carlo Beenakker, Markus B\"uttiker, and Yuli Nazarov for valuable
discussions. This research was supported by the ``Ne\-der\-land\-se
or\-ga\-ni\-sa\-tie voor We\-ten\-schap\-pe\-lijk On\-der\-zoek'' (NWO) and by
the ``Stich\-ting voor Fun\-da\-men\-teel On\-der\-zoek der Ma\-te\-rie''
(FOM), by the Swiss National Science Foundation, and by the European
Community's Human Potential Programme for Nanoscale Dynamics, Coherence and
Computation under contract HPRN-CT-2000-00144.
\end{acknowledgements}


\begin{thebibliography}{99}
\bibitem{Earlier1}
This type of energy transfer statistics has been used in Ref.
\cite{Pil03} for open point contacts to calculate the charge
noise of a chaotic cavity in the hot electron regime.
\end{thebibliography}
\end{document}